\documentclass[12pt,a4paper]{article}

\usepackage{amsmath}

\title{Dark Energy Accretion onto a Black Holes in an Expanding Universe}
\author{\textbf{Cheng-Yi Sun\footnote{cysun@mails.gucas.ac.cn; ddscy@163.com}\ $^{,a}$\
}\\ \\
 {$^a$\small Institute of Modern Physics, Northwest University,}\\
     \small Xian 710069, P.R. China.}

\begin{document}
\maketitle
\begin{abstract}
By using the solution describing a black hole embedded in the FLRW
universe, we obtain the evolving equation of the black hole mass
expressed in terms of the cosmological parameters. The evolving
equation indicates that in the phantom dark energy universe the
black hole mass becomes zero before the Big Rip is reached.
\end{abstract}
\section{Introduction}
In recent years, strong observational evidence shows that the
current expansion of our universe is accelerating \cite{supernova}.
In the framework of the general relativity, this positive
acceleration means that, at present, our universe is dominated by a
mysterious component, the dark energy \cite{t0603057}. Usually, we
may model the dark energy by a perfect fluid with a negative
pressure $p=w\rho<0$. The dark energy may be in the form of a vacuum
energy with $w=1$, or a quintessence model with $w>-1$. The
observation gives the constraint \cite{a0701047}:
$w=-1.07^{+0.09}_{-0.09}$(stat $1\sigma$)$\pm0.13$(sys). Then there
exists the possibility of the phantom models with $w<-1$. In this
case, the universe undergoes the ``supper accelerated" expansion.
Finally, in the finite time, the Big Rip singularity is reached, and
all bounded objects in the universe are torn apart \cite{Caldwell}.

What would happen to black holes in an expanding universe filled
with the phantom dark energy? In \cite{g0402089,a0505618}, using the
Schwarzschild metric,  the authors have obtained the result
\begin{equation}
  \label{dmdt0}
  \dot{M}\equiv\frac{dM}{dt}=4\pi G^2AM^2[p_\infty+\rho_\infty],
\end{equation}
where $p_\infty$ and $\rho_\infty$ are respectively the pressure and
energy density of the universe at the asymptotic limit
$r\rightarrow+\infty$. And $A$ is a dimensionless constant. Then the
author concluded that the masses of all black holes will tend to
vanish as the universe approached the Big Rip. In \cite{a0603761},
the authors obtained the same result (\ref{dmdt0}), by using the
non-static Schwarzschild metric
\begin{equation}
  \label{nonStaticBLMetric}
  ds^2=-\left(1-\frac{2GM(t)}{r}\right)dt^2+\left(1-\frac{2GM(t)}{r}\right)^{-1}dr^2+r^2(d\theta^2+\sin^2\theta
  d\phi^2).
\end{equation}
However, the metric is asymptotically flat and does not describe
exactly the space-time of a black hole embedded in the FLRW
universe. Actually, Eq.(\ref{dmdt0}) is obtained by ignoring the
backreaction of the phantom matter on the black hole. When the
matter density of the background is low, the ignorance is
reasonable. However, as the Big Rip is approached, the matter
density becomes very large. The backreaction should not be
neglected, and the metric (\ref{nonStaticBLMetric}) becomes
unsuitable for describing the spacetime. Then, we should adopt some
exact metric, in order to take into account of the backreaction.

Recently, new exact solutions describing black holes embedded in the
FLRW universe have been suggested in \cite{0707.1350,0802.1298}.
Then it is interesting for us to study the problem of the dark
energy accretion on black holes by using the exact solutions. In
\cite{0707.1350}, the accretion rate (Hawking-Hayward quasi-local
mass added per unit time) is obtained. But the result is not
suitable for the application in the cosmology. In order to study the
evolution of the black hole mass as the universe expands, we should
express the accretion rate in terms of the parameters at the
asymptotic limit $r\rightarrow+\infty$, namely the cosmological
parameters, $p_\infty,\ \rho_\infty$, et al..

In this paper we will try to obtain such an expression by using the
solution with imperfect fluid and radial mass flow in
\cite{0707.1350}. Our result shows that the evolution of the black
hole mass is governed by a second-order differential equation, which
is different from Eq.(\ref{dmdt0}).

The paper is organized as follows. In Section \ref{Recall}, we
recall the results in \cite{0707.1350}. In Section \ref{Finding}, we
show the evolving equation of the black hole mass expressed in terms
of the cosmological parameters. Section \ref{Discussion} contains a
discussion and conclusions.

\section{the Solution with Imperfect Fluid and Radial Mass Flow}
\label{Recall}

In \cite{0707.1350}, the line element is written in the form
\begin{equation}
  \label{line element}
  ds^2=-\frac{B^2(t,r)}{A^2(t,r)}dt^2+a^2(t)A^4(t,r)(dr^2+r^2d\Omega^2),
\end{equation}
with $A(t,r)=1+\frac{Gm(t)}{2r}$ and $B(t,r)=1-\frac{Gm(t)}{2r}$.
Here and after, we take $c=\hbar=1$. Then, at the asymptotic limit
$r\rightarrow+\infty$, the metric becomes the FLRW metric. In this
space-time, the physically relevant mass is the Hawking-Hayward
quasi-local mass \cite{0707.1350,Hawking,Hayward}
\begin{equation}
  \label{Hawking-Hayward mss}
  m_H(t)=a(t)m(t).
\end{equation}

In this paper, we study the solution with radial mass flow,
$u(r,t)$. The cosmological matter is assumed to be described by the
imperfect fluid energy-momentum tensor
\begin{equation}
  \label{MEtensor}
  T^{ab}=(p+\rho)u^au^b+pg^{ab}+q^au^b+q^bu^a,
\end{equation}
with
\begin{equation}
  \label{uAndq}
  u^a=\left(\frac{A}{B}\sqrt{1+a^2A^4u^2},u,0,0\right),\ \ \ \
  q^a=(0,q,0,0).
\end{equation}
Here, the purely spatial vector field $q^a$ describes a radial
energy flow. The condition
\[
\lim_{r\rightarrow+\infty}{u(t,r)}=\lim_{r\rightarrow+\infty}{q(t,r)}=0,
\]
should be imposed, in order for the momentum-energy tensor
(\ref{MEtensor}) to become the perfect fluid momentum-energy tensor
in cosmology at the asymptotic infinity $r\rightarrow+\infty$,
\[
  T_\infty^{ab}=(p_\infty+\rho_\infty)u_\infty^au_\infty^b+p_\infty g_\infty^{ab}.
\]

The Einstein equations give four independent equations
\cite{0707.1350},
\begin{eqnarray}
  \label{01dmHdt}
  \dot{m}_H&=&-\frac{1}{2}aB^2\mathcal{A}\sqrt{1+a^2A^4u^2}(p+\rho)u,\\
  \label{p}
  8\pi Gp&=&-\left(\frac{A}{B}\right)^2\left[3C^2+2(\dot{C}+\frac{G\dot{m}C}{rAB})\right],\\
  \label{rho}
  8\pi
  G\rho&=&\left(\frac{A}{B}\right)^2\left[3C^2+(\dot{C}+\frac{G\dot{m}C}{rAB})\frac{2a^2A^4u^2}{1+a^2A^4u^2}\right],\\
  \label{q}
  q&=&-(p+\rho)\frac{u}{2},
\end{eqnarray}
with $\mathcal{A}\equiv \int\int d\theta
d\varphi\sqrt{g_\Sigma}=4\pi r^2a^2A^4$ representing the area of a
spherical surface with isotropic radius $r$, and
$C\equiv\frac{\dot{a}}{a}+\frac{2\dot{A}}{A}$. Here and after, the
dot denotes the derivative with respect to the time.

This is the solution describing a black hole embedded in the FLRW
universe \cite{0707.1350}. As $r\rightarrow\infty$, Eq.(\ref{rho})
becomes the Friedmann equation
\[
  8\pi G\rho=3\frac{\dot{a}^2}{a^2}.
\]
Then Eq.(\ref{rho}) may be consider as a generalized Friedmann
equation. Eq.(\ref{p}) and (\ref{rho}) indicate a generalized
conservation law
\begin{equation}
  \label{conservationLaw1}
  \begin{split}
    \dot{\rho}(1+a^2A^4u^2)+\dot{p}(a^2A^4u^2)
    &+3(\frac{\dot{a}}{a}+2\frac{\dot{A}}{A})(p+\rho)(1+a^2A^4u^2)\\
      &+(p+\rho)2a^2A^4u^2(\frac{\dot{a}}{a}+2\frac{\dot{A}}{A}+\frac{\dot{u}}{u})=0,
  \end{split}
\end{equation}
which, as $r\rightarrow+\infty$, becomes the conservation law in the
FLRW universe
\[
  \dot{\rho}+3\frac{\dot{a}}{a}(p+\rho)=0.
\]
Additionally, the radial component of the energy-momentum
conservation law, ${T_1}^{\mu}_{;\mu}=0$, gives another conservation
law
\begin{equation}
  \label{conservationLaw2}
  \begin{split}
    0&=\frac{a^2A^5}{2B\sqrt{1+a^2A^4u^2}}\{5(\frac{\dot{a}}{a}+2\frac{\dot{A}}{A})(1+a^2A^4u^2)(p+\rho)u+(p+\rho)a^2A^4u^3
                       (\frac{\dot{a}}{a}+2\frac{\dot{A}}{A}+\frac{\dot{u}}{u})\\
     &\ \
       +(1+a^2A^4u^2)[(\dot{p}+\dot{\rho})u+(p+\rho)\dot{u}]\}+p'+(\frac{B'}{B}-\frac{A'}{A})(p+\rho)(1+a^2A^4u^2)
  \end{split}
\end{equation}
Here, we note that the time component of the energy-momentum
conservation law, ${T_0}^{\mu}_{;\mu}=0$, does not give an
independent equation, which can be derived by using the equations
(\ref{01dmHdt}) and (\ref{conservationLaw1}).

\section{the Evolving Equation in terms of Cosmological Parameters}
\label{Finding}

Eq.(\ref{01dmHdt}) gives the accretion rate. Since $u<0$, the mass
$m_H$ increases if $p+\rho>0$, stays constant in a de-Sitter
background, and decreases for the phantom dark energy $p+\rho<0$.
However, this equation is not convenient for us to study the
evolution of the black hole mass as the universe expands. In
cosmology, what we need is an evolving equation of $m_H$ expressed
in terms of the parameters at the asymptotic limit
$r\rightarrow+\infty$, namely the cosmological parameters.

To obtain such an evolving equation, we may take the limit of
Eq.(\ref{01dmHdt}) as $r\rightarrow+\infty$. Then we have
\begin{equation}
  \label{01dmHdtLimit}
  \dot{m}_H
  =-2\pi a^3(p_\infty+\rho_\infty)\lim_{r\rightarrow+\infty}{(
  ur^2)},
\end{equation}
where $p_\infty(t)$ and $\rho_\infty(t)$ are respectively the limit
of $p(t,r)$ and $\rho(t,r)$ as $r\rightarrow+\infty$. This result
indicates that, as $r\rightarrow+\infty$,
\begin{equation}
  \label{uInfty}
  u(t,r)\simeq\frac{u_\infty(t)}{r^2}+\mathcal{O}(r^{-3}).
\end{equation}
Similarly, as $r\rightarrow\infty$, we may expect
\begin{eqnarray}
  \label{pInfty}
  p(t,r)\simeq p_\infty(t)+\frac{p_1(t)}{r}+\mathcal{O}(r^{-2}),\\
  \label{rhoInfty}
  \rho(t,r)\simeq
  \rho_\infty(t)+\frac{\rho_1(t)}{r}+\mathcal{O}(r^{-2}).
\end{eqnarray}
Then, up to the order $r^{-2}$, Eq.(\ref{conservationLaw2}) gives
\begin{equation}
  \label{limitOfradialCL}
  \frac{d}{dt}[a^5(p_\infty+\rho_\infty)u_\infty]=2a^3[p_1-Gm(p_\infty+\rho_\infty)]
\end{equation}
At the same time, up to the order $r^{-1}$,
Eq.(\ref{conservationLaw1}) indicates two equations
\begin{eqnarray}
  \label{FLRWCL}
  \dot{\rho}_\infty&+&3\frac{\dot{a}}{a}(p_\infty+\rho_\infty)=0,\\
  \label{firstOrderOfCL1}
  \dot{\rho}_1&+&3\frac{\dot{a}}{a}(p_1+\rho_1)+3G\dot{m}(p_\infty+\rho_\infty)=0.
\end{eqnarray}
Here, naturally, we can impose the equation of state
\begin{equation}
  \label{EOS}
  p_\infty(t)=w\rho_\infty(t),
\end{equation}
where $w$ is a dimensionless constant.  Further, we assume the same
equation of state for $p_1(t)$ and $\rho_1(t)$
\begin{equation}
  \label{EOS1}
  p_1(t)=w\rho_1(t).
\end{equation}
Together the two equations and Eq.(\ref{FLRWCL}), we can solve
Eq.(\ref{firstOrderOfCL1}), and obtain
\begin{equation}
  \label{rho-1}
  \rho_1(t)=3G(m_0-m)(p_\infty+\rho_\infty),
\end{equation}
where $m_0$ is an integral constant with the dimension of mass. Now,
Eq.(\ref{limitOfradialCL}) can be rewritten as
\begin{equation}
  \label{dUdt}
  \frac{d}{dt}[a^5(p_\infty+\rho_\infty)u_\infty]=2Ga^3[3wm_0-(3w+1)m](p_\infty+\rho_\infty).
\end{equation}
Using this equation and Eq.(\ref{01dmHdtLimit}), finally we can
obtain the evolving equation of $m_H$
\begin{equation}
  \label{dmHddt}
  \ddot{m}_H+2\frac{\dot{a}}{a}\dot{m}_H-4\pi
  G[(3w+1)m_H-3wm_0a](p_\infty+\rho_\infty)=0,
\end{equation}
where $\ddot{m}_H\equiv \frac{d^2m_H}{dt^2}$. This is a second-order
differential equation as we expect from the second-order
differential equations (\ref{01dmHdt})-(\ref{rho}).

\section{Discussion and Conclusion}
\label{Discussion}

In the last section, we obtain the evolving equation of the mass of
a black hole embedded in the FLRW universe. Both Eq.(\ref{dmdt0})
and our result, Eq.(\ref{dmHddt}), imply that the black hole mass
decrease in the phantom dark energy universe, but the qualitatively
evolving behaviors of the black hole mass indicated by the two
equation are different. Due to Eq.(\ref{dmdt0}), it has been shown
in \cite{g0402089,0806.1080} that the mass of a black hole tends to
zero in the phantom dark energy universe approaching the Big Rip.
However, by using our result Eq.(\ref{dmHddt}), we find that the
black hole mass in the dark energy universe has been zero before the
Big Rip singularity is reached.

In order to show this, firstly let us rewrite Eq.(\ref{dmHddt}) in
terms of the derivative with respect to the scale factor $a(t)$,
\begin{equation}
  \label{dmHdda}
  \frac{d^2m_H}{da^2}+\frac{3(1-w)}{2}a^{-1}\frac{dm_H}{da}=\frac{3}{2}(3w+1)(1+w)a^{-2}m_H-\frac{9}{2}w(1+w)m_0a^{-1}.
\end{equation}
Here, we have used the equations
\[
  \left(\frac{\dot{a}}{a}\right)^2=\frac{8\pi G}{3}\rho_\infty,\ \
  \rho_\infty\propto a^{-3(1+w)}.
\]
The general solution of Eq.(\ref{dmHdda}) is
\begin{equation}
  \label{mH}
  m_H(t)=C_1[a(t)]^{1+3w}-C_2[a(t)]^{-3(1+w)/2}+\frac{3(1+w)}{3w+5}m_0a(t),
\end{equation}
where $C_1$ and $C_2$ are the integral constants. Use $t_0$ to
denote the initial moment. The two constants are determined by the
initial mass, $m_H(t_0)$, and the initial accretion rate
$\dot{m}_H(t_0)$.

In a phantom dark energy scenario, $w$ is close to $-1$ and $w<-1$.
Then we may have
\[
1+3w<0,\ \ 1>-3(1+w)/2>0,\ \ \frac{3(1+w)}{3w+5}<0.
\]
Using $m_H(t_0)>0$ and $\dot{m}_H(t_0)<0$, generally we may have
\[
  C_1>0,\ \ C_2>0.
\]
Then Eq.(\ref{mH}) indicates that, before the Big Rip singularity is
reached, the black mass has been zero at the moment $t_e$ which is
determined approximately by the equation
\[
  C_1[a(t_e)]^{1+3w}=C_2[a(t_e)]^{-3(1+w)/2}-\frac{3(1+w)}{3w+5}m_0a(t_e).
\]
Physically, this result seems to be better, since the Big Rip
singularity becomes unnecessary for the disappearance of the black
holes. Of course, the analysis above may be not rigid when the black
hole mass is at the order of the Planck mass, since the effect of
the Hawking radiation might dominate the evolution of the black hole
with the Planck mass.

In \cite{a0603761}, the author have shown that Eq.(\ref{dmdt0}) can
be generalized to the non-static metric (\ref{nonStaticBLMetric}).
Then we might expect that Eq.(\ref{dmdt0}) should be reduced by
taking $a(t)$ to be constant, $\dot{a}=0$, in Eq.(\ref{dmHddt}).
However, it is not the case. The reason is that Eq.(\ref{dmdt0}) in
\cite{a0603761} is obtained under the assumption
\[
  \lim_{r\rightarrow\infty}(ur^2)=AG^2M^2.
\]
From Eq.(\ref{limitOfradialCL}), we know the assumption for the
solution used in this paper is unreasonable. Then it becomes natural
that Eq.(\ref{dmdt0}) can not be reduced from Eq.(\ref{dmHddt}) with
$\dot{a}=0$.

Even, we should note that the previously known solution for the
perfect fluid, Eq.(\ref{dmdt0}), cannot be reduced from our result
Eq.(\ref{dmHddt}) by taking $q^a=0$ in Eq.(\ref{MEtensor}). In fact,
$q^a\neq0$ is necessary. It has been shown in \cite{0707.1350} that,
for the perfect fluid described by taking $q^a=0$ in
Eq.(\ref{MEtensor}), the Einstein equations indicate
\[
  p+\rho=0.
\]
This can be also learned from Eq.(\ref{q}). This implies that only
the de-Sitter equation of state $p=-\rho$ is allowed.
Eq.(\ref{01dmHdt}) accordingly yields $\dot{m}_H=0$. So no accretion
happens for a single perfect fluid.  Then the imperfect fluid
(\ref{MEtensor}) with $q^a\neq0$ is necessary for the accretion to
happen when the metric (\ref{line element}) is used. This indicates
that the previously known accretion rate (\ref{dmdt0}) is not the
limit of our result (\ref{dmHddt}) as $q^a\rightarrow0$.

However, We do not think our result (\ref{dmHddt}) is better than
the result (\ref{dmdt0}). The reason is that the imperfect fluid
solution which our result (\ref{dmHddt}) is based on has the
unphysical behavior \cite{0707.1350}---the superluminal motion of
the fluid as $r\rightarrow m/2$. For this reason£¬our result cannot
be used as a "better approximation" as compared with the result
(\ref{dmdt0}). We may take Eq.(\ref{dmHddt}) as an interesting try.

Summarily, in the paper, by using the exact solution describing a
black hole embedded in the FLRW universe \cite{0707.1350} and the
equation of state (\ref{EOS}), we obtain the evolving equation of
the black hole mass expressed in terms of the cosmological
parameters, which is convenient for the application in cosmology.
Our result,  a second-order differential equation (\ref{dmHddt}), is
different from the result (\ref{dmdt0}) in \cite{g0402089}. Our
result indicates that the black mass in the phantom dark energy
universe has been zero at the moment $t_e$ before the Big Rip moment
$t_B$, while, due to Eq.(\ref{dmdt0}), the mass of a black hole
tends to zero in the phantom dark energy universe approaching the
Big Rip. Here, we emphasize again that our result (\ref{dmHddt}) is
not better than the result (\ref{dmdt0}) because of the unphysical
behavior of the imperfect fluid solution which our result
(\ref{dmHddt}) is based on.


\begin{thebibliography}{99}

\bibitem{supernova}A. G. Riess et al. Astron. J. 116, 1009 (1998) [astro-ph/9805201]; S. Perlmutter et al. Astrophys. J. 517, 565
(1999) [astro-ph/9812133]; J. L. Tonry et al. Astrophys. J. 594, 1
(2003) [astro-ph/0305008]; R. A. Knop et al. Astrophys. J. 598, 102
(2003) [astro-ph/0309368]; A. G. Riess et al. Astrophys. J. 607, 665
(2004) [astro-ph/0402512].

\bibitem{t0603057}E. J. Copeland, M. Sami and S. Tsujikawa,
Int. J. Mod. Phys. D 15, 1753  (2006); [arXiv:hep-th/0603057].

\bibitem{a0701047}W. M. Wood-Vasey, \emph{et al.},
Astrophys. J. 666, 694 (2007); [arXiv:astro-ph/0701047].

\bibitem{Caldwell}R. Caldwell, M. Kamionkowski, N. Weinberg, Phys. Rev. Lett.
91, 071301 (2003); R. Caldwell,
Phys. Lett. B 545, 23 (2002), [arXiv:astro-ph/9908168].

\bibitem{g0402089}E. Babichev, V. Dokuchaev and Yu. Eroshenko,
Phys. Rev. Lett. 93, 021102 (2004); [gr-qc/0402089].

\bibitem{a0505618}E. Babichev, V. Dokuchaev, Yu. Eroshenko,
J. Exp. Theor. Phys. 100, 528 (2005); [arXiv:astro-ph/0505618].

\bibitem{a0603761}P. Martin-Moruno, J. A. J. Madrid and P. F. Gonzalez-Diaz,
Phys. Lett. B 640, 117 (2006) [arXiv:astro-ph/0603761].


\bibitem{0707.1350}V. Faraoni and A. Jacques,
Phys. Rev. D 76, 063510 (2007); arXiv:0707.1350[gr-qc].


\bibitem{0802.1298}C. Gao, X. Chen, V. Faraoni and Y. G. Shen,
Phys. Rev. D 78, 024008 (2008); arXiv:0802.1298[gr-qc].

\bibitem{Hawking}S.W. Hawking, J. Math. Phys. 9, 598 (1968).

\bibitem{Hayward}S.A. Hayward,
 Phys. Rev. D 49, 831 (1994); [arXiv:gr-qc/9303030]

\bibitem{0806.1080}C.Y. Sun,
Phys. Rev. D 78, 064060 (2008); arXiv:0806.1080[hep-th].


\end{thebibliography}
\end{document}